\documentstyle[aps,epsf,prc]{revtex}

\begin{document}
\title{Threshold $J/\psi-$ production in nucleon-nucleon collisions}
\author{Michail P. Rekalo \footnote{Permanent address:
\it National Science Center KFTI, 310108 Kharkov, Ukraine}
}
\address{Middle East Technical University, 
Physics Department, Ankara 06531, Turkey}
\author{Egle Tomasi-Gustafsson}
\address{\it DAPNIA/SPhN, CEA/Saclay, 91191 Gif-sur-Yvette Cedex, 
France}
\date{\today}

\maketitle
\begin{abstract}
We analyze $J/\psi-$ production in nucleon-nucleon collisions near threshold in the framework of a general model independent formalism, which can be applied to any reaction $N+N\to N+N+V^0$, where $V^0=\omega$, $\phi$, or $J/\psi$. Such reactions show large isotopic effects: a large difference for $pp$- and $pn$-collisions, which is due to the different spin structure of the corresponding matrix elements. The analysis of the spin structure and of the polarization observables is based on symmetry properties of the strong interaction. Using existing experimental data on the different decays of $J/\psi-$meson, we suggest a model for $N+N\to N+N+J/\psi$, based on $t-$channel $\eta+\pi$-exchanges. We predict polarization phenomena for the $n+p\to n+p+J/\psi$-reaction and the ratio of cross sections for $np$ and $pp$-collisions. For the processes $\eta(\pi)+N\to N+J/\psi$ we apply two different approaches: vector meson exchange and local four-particle interaction. In both cases we find larger $J/\psi$-production in $np$-collisions, with respect to $pp$-collisions.

\end{abstract}
\section{Introduction}
It is well known that the $J/\psi-$meson has been observed in two different reactions: in $p+Be$-collisions\cite{Aub74} and $ e^+e^-$ \cite{Aug74}.

Since that time  experimental and theoretical studies of $J/\psi-$production  have been going on. As a result of high statistics and high resolution experiments a large information on the properties of the $J/\psi-$meson have been collected, on the production processes and on its numerous decays. From a theoretical point of view, the interpretation of the data, in particular in confinement regime, is very controversial. As an example, the $c-$quark mass is too large, if compared to predictions from chiral symmetry, but for theories based on expansion of heavy quark mass (Heavy Quark Effective Theory), this mass is too small \cite{Is89}. In principle all data on $J/\psi-$ should be explained in the framework of QCD. For example, the typical picture of charm production (open or hidden) should unify the different steps: parton-level hard process with production of $c\overline{c}$-pairs, after the step of hadronization of
$c\overline{c}$- into $J/\psi$ or into charmed hadrons (mesons and baryons) and the final state interaction of the produced charmed hadrons with other particles. The relatively large transferred momenta involved in most processes of $J/\psi-$production in hadron-hadron collisions allow to treat the first step in framework of perturbative QCD. But the applicability of QCD is not so straightforward for the description of the $c$-quark hadronization.

This is correct especially for exclusive reactions, such as, for example, $J/\psi$ production in nucleon-nucleon collision:  $N+N\to N+N+J/\psi$. Note that the threshold for this reaction is large: $E_{th}$=12.24 GeV which corresponds to $\sqrt{s}=2m+m_{J/\psi}\simeq$ 5 GeV, whereas  experimental data  about $p+p\to J/\psi + X$ are available for $\sqrt{s} \ge$ 6.7 GeV \cite{Ba78}. therefore the experimental study of  $N+N\to N+N+J/\psi$ near threshold is an important part of the program of the future accelerator facility at GSI \cite{GSI}.

Precise measurements of threshold $J/\psi$ production will bring important information with respect to the following issues. First of all the threshold meson production in $NN$-collisions gives deeper insight in the reaction mechanisms as it is shown by recent experiments on $p+p\to p+p + \omega(\phi)$ \cite{Bo68,Co67,Ca68,Co70,Ye70,Al68,Hi99,Ab01}, $p+p\to \Lambda(\Sigma^0)+K^+ +p $ \cite{Ba98,Bi98,Se98,Ba99}, and $p+p\to p+p + \eta(\eta ')$ \cite{Bo68,Co67,Ca68,Co70,Ye70,Al68,Sm99,Ca96,Be93,Ch94,Pi62,Al68,Hi98,Mo00,Ba00}. In this respect, $J/\psi-$production has a specific interest: the production and the propagation of charm in ion-ion collisions has been considered as one of the most promising probe of quark-gluon plasma (QGP) formation \cite{cern}. The  productions of charmonium (hidden strangeness) and $D$ $(D^*)$ mesons (open charm) are equally important. The suppression of charmonium production in heavy ion collisions has been indicated as a signature of QGP \cite{Ma86}, but in order to state the evidence of a clear signal, it is necessary to analyze in detail all possible mechanisms for $J/\psi-$production in ion-ion collisions, and also all other processes which are responsible for the dissociation of the produced $J/\psi-$meson,  such as $J/\psi+N\to \Lambda_c+D$, for example.

The aim of this paper is to study $J/\psi-$production in the simplest $NN$-reactions: $p+p\to p+p+J/\psi$ and  $n+p\to n+p+J/\psi$. In the near threshold region there is no experimental information about a possible difference in $pp$- and $np$-collisions. There are no theoretical predictions for such exclusive processes, which are very difficult to treat in framework of QCD. It is possible to find some phenomenological parametrizations \cite{Si97} of the energy dependence of the total cross section for $p+N\to J/\psi+X$, without any consideration on possible isotopic effects. However, previous experience   with light meson production \cite{Ti01}, $N+N\to N+N + \omega(\phi)$, showed an essential difference in $np-$ and $pp-$cross sections, in particular in the near threshold region, and there is no physical reason for a different behavior in $J/\psi-$production. It is evident that the knowledge of the elementary process  $p+N\to p+N+J/\psi$ is very important for a realistic calculation of $J/\psi-$production in nucleus-nucleus collisions. 

In principle the 'elastic' $J/\psi-$production in $NN$-collisions can be treated in full analogy with processes of light vector meson production. All symmetry properties of the strong interaction, such as the Pauli principle, the isotopic invariance, the P-invariance, which have been successfully applied to light vector meson production in $NN$-collisions \cite{Re97}, hold for $J/\psi-$production, too. A formalism can be built, which is particularly simplified in the threshold region, where final particles are produced in $S$-state. Simple considerations indicate that this region is quite wide: the effective proton size, which is responsible for charm creation, has to be quite small, $r_c\simeq 1/m_c$, where $m_c$ is the $c$-quark mass  \cite{Br01}. Therefore the $S$-wave picture can be applied for $q\le m_c$, where $q$ is the $J/\psi-$ three-momentum in the reaction center of mass (CMS). 

This paper is organized as follows. In the first chapter we establish the spin structure of the threshold matrix for the two $NN$-processes:
$$p+p\to p+p+J/\psi,$$
$$p+n\to p+n+J/\psi$$
in terms of partial $S-$wave amplitudes, and calculate the simplest polarization observables in  terms of these amplitudes. In Chapter 2 we treat the dynamical issue in terms of $t-$channel exchange mechanisms by light mesons. We give numerical predictions in framework of a $\pi+\eta$-model. The experimental data about different hadronic decays of $J/\psi-$meson may give constrains on our predictions.
\section{Spin structure of threshold matrix elements and polarization phenomena}
In the general case the spin structure of the matrix element for the process 
$N+N\to N+N+V$ is described by a set of 48 independent complex amplitudes, which are functions of five kinematical variables. The same reaction, in coplanar kinematics, is described by 24 amplitudes, functions of four variables. In collinear kinematics the number of independent amplitudes is reduced to 7 and the description of this reaction is further simplified in case of threshold $V$-meson production, where all final particles are in $S$-state.

Applying the selection rules following from the Pauli principle, the P-invariance and the conservation of the total angular momentum, it is possible to prove that the threshold process $p+p \rightarrow p+p+V^0$ is characterized by a single partial transition: 
\begin{equation}
S_i=1,~\ell_i=1~\to ~{\cal J}^{ P}=1^- \to S_f=0, 
\label{eq:eqpp}
\end{equation}
where $S_i$ ($S_{f}$) is the total spin of the two protons in the initial (final) 
states and $\ell_i$ is the orbital momentum of the colliding protons.
In the CMS of the considered reaction, the matrix element corresponding to transition (\ref{eq:eqpp}) can be written as:
\begin{equation}
{\cal M}(pp)=2f_{10}(\tilde{\chi}_2~\sigma_y ~\vec{\sigma}
\cdot\vec  U^* \times\hat{\vec k }\chi_1)~(\chi^{\dagger}_4 \sigma_y\ \tilde{\chi}^{\dagger}_3 )
\label{eq:mpp},
\end{equation}
where $\chi_1$ and $\chi_2$ ( $\chi_3$ and $\chi_4$) are the
two-component spinors of the initial (final) protons;  $\vec  U$ is the three-vector of the $V$-meson polarization, $\hat{\vec k}$ is
the unit vector along the 3-momentum of the initial proton; $f_{10}$ is the S-wave partial amplitude, describing the triplet-singlet transition of the two-proton system in V-meson production.

In case of  $np$-collisions, applying the conservation of isotopic invariance for the strong interaction, two threshold partial transitions are allowed:
$$S_i=1,~\ell_i=1~\to ~{\cal J}^{ P}=1^- \to S_f=0,$$
\begin{equation}
S_i=0,~\ell_i=1~\to ~{\cal J}^{ P}=1^- \to S_f=1,
\label{eq:eqnp}
\end{equation}
with the following spin structure of the matrix element:
\begin{eqnarray}
&{\cal M}(np)=&f_{10}(\tilde{\chi}_2~\sigma_y ~\vec{\sigma}
\cdot\vec  U^* \times\hat{\vec k }\chi_1)~(\chi^{\dagger}_4 \sigma_y\ \tilde{\chi}^{\dagger}_3 )+\nonumber \\
&& f_{01}(\tilde{\chi}_2~\sigma_y\chi_1)(\chi^{\dagger}_4
\vec{\sigma}
\cdot\vec  U^* \times\hat{\vec k }\sigma_y \tilde{\chi}^{\dagger}_3 ),
\label{eq:mnp}
\end{eqnarray}
where $f_{01}$ is the S-wave partial amplitude describing the singlet-triplet transition of the two-nucleon system in V-meson production. In the general case the amplitudes $f_{10}$ and $f_{01}$ are complex functions, depending on the energies $E$, $E'$ and $E_V$, where $E,(E')$ and $E_V$ are the energies of the initial (final) proton and of the produced $V-$meson, respectively.

Note that $f_{10}$ is the common amplitude for $pp$- and $np$-collisions, due to the isotopic invariance of the strong interaction. This explains the presence of the coefficient 2 in Eq. (\ref{eq:mpp}).
The parametrizations (\ref{eq:mpp}) and (\ref{eq:mnp}) are model independent descriptions of the spin structure for threshold production of any vector meson in $NN$-collisions, $N+N\to N+N+V$, from the light $\rho$, $\omega$ and $\phi$ to $J/\psi$, 
$\psi'$,$\psi''$, including the vector bottonium: $\Upsilon(1S)$, $\Upsilon(2S)$ and even the hypothetical exotic vector toponium  states. All dynamical information is contained in the partial amplitudes $f_{01}$ and $f_{10}$, which are different for the different vector particles. On the opposite, some polarization phenomena have common characteristics, essentially independent from the type of vector meson. For example, vector mesons produced in $pp$- and $np$-threshold collisions are transversally polarized,
and the elements of the density matrix $\rho$ are independent from the relative values of the amplitudes $f_{01}$ and $f_{10}$: $\rho_{xx}=\rho_{yy}=\frac
{1}{2}$, $\rho_{zz}=0$. Therefore, the angular distribution shows the  $\sin^2\theta_P$-dependence for the subsequent decay $V^0\to P+P$ (where $P$ is a pseudoscalar meson) and the $(1+\cos^2\theta)$-dependence  for the decay
$V^0\to \mu^+ +\mu^-$, where $\theta$ ($\theta_P$) is the angle between $\hat{\vec k}$ and the $\mu^-$ $(P)$ momentum (in the rest system of $V^0$). Possible deviations from this behavior have to be considered as an indication of the presence of higher partial waves in the final state. 

All other one-spin polarization observables, related to the polarizations of the initial or final nucleons, identically vanish, for any process of $V-$meson production.

The dependence of the differential cross section for threshold collisions of polarized nucleons (where the polarization of the final particles is not detected) can be parametrized as follows:

\begin{equation}
\displaystyle\frac{d\sigma}{d\omega}(\vec P_1,\vec P_2)=\left ( \displaystyle\frac{d\sigma}{d\omega}\right)_0
\left (1+{\cal A}_1 \vec P_1 \cdot \vec P_2 +{\cal A}_2 \hat{\vec k} \cdot\vec P_1\hat{\vec k} \cdot\vec P_2 \right ),
\label{eq:sig}
\end{equation}
where $\vec P_1 $ and $\vec P_2$ are the axial vectors of the beam and target nucleon polarizations, $d\omega$ is the element of phase-space for the three-particle final state. The spin correlation  coefficients  ${\cal A}_1$ and ${\cal A}_2$ are real and they are different for $pp$- and $np$- collisions:
\begin{itemize}
\item $\vec p+\vec p\to p+p+V^0$:  ${\cal A}_1(pp)=0$, ${\cal A}_2(pp)=1$.
\item $\vec n+\vec p\to n+p+V^0$:  
${\cal A}_1(np)=-\displaystyle\frac{|f_{01}|^2}{|f_{01}|^2+|f_{10}|^2},~~{\cal
A}_2(np)=\displaystyle\frac{|f_{10}|^2}{|f_{01}|^2+|f_{10}|^2}$,
\end{itemize}
with the following relations $-{\cal A}_1(np)+{\cal A}_2(np)=1$ and 
$0\le {\cal A}_2(np)\le 1$.

Defining ${\cal R}$ as the ratio of the total (unpolarized) cross section for $np$- and $pp$- collisions, taking into account the identity of final particles in $p+p\to p+p+V^0$, we find:
\begin{equation}
{\cal R}=\displaystyle\frac{\sigma(np\to npV^0)}
{\sigma(pp\to ppV^0)}=\displaystyle\frac{1}{2}+\displaystyle\frac{1}{2}
\displaystyle\frac{|f_{01}|^2}{|f_{10}|^2}.
\label{eq:eqr}
\end{equation}
So the following relation holds:
$$
{\cal A}_1=-1+\displaystyle\frac{1}{2{\cal R}}.
$$

The polarization transfer from the initial neutron to the final proton 
($\vec n+p\to n+\vec p+V$), can be parameterized as follows:
\begin{equation}
{\cal P}_f=p_1 \vec P_1+p_2 \hat{\vec k} (\hat{\vec k}\cdot\vec P_1),
\label{eq:ptr}
\end{equation}
with a simple expression, which relates the real coefficients $p_1$ and $p_2$ to the partial amplitudes $f_{01}$ and $f_{10}$:
$$p_1(np) =-p_2(np)=\displaystyle\frac{2{\cal R}e f_{01}f_{10}^*}{|f_{01}|^2+|f_{10}|^2}=\cos\delta\displaystyle\frac
{\sqrt{2{\cal R}-1}}{{\cal R}},$$
where $\delta$ is the relative phase of $f_{01}$ and $f_{10}$, which is non zero, in the general case. 

For the process $p+p\to p+p+V^0$ the relation $p_1(pp) =p_2(pp)=0$ holds, for any vector meson $V^0$.

\section{The dynamics for the $\lowercase{t}$-channel}

The parameterization of the spin structure of the threshold matrix elements given above, is based on fundamental symmetry properties. It is therefore model independent and can be applied to any reaction mechanism. Following the standard way in describing the nucleon-nucleon interaction, we will apply $t-$channel $\pi^0$, $\eta$, $\sigma$ and $\rho (\omega)$  meson exchanges to $J/\psi$ production, too. Such approach has been used to describe the production of light vector mesons such as $\phi$ and  $\omega$ \cite{Si97,Ti01}. The reaction threshold for $p+p\to p+p+J/\psi$ in the laboratory system (LAB) is quite large. However, the formalism of Pomeron exchange can not be applied here, even at such large energies, because the Regge picture is valid when not only the initial energy is large, but also the excitation energy: the quantity $(W-W_{th})/W_{th}$ (where $W$ is the total energy) has to be essentially larger than unity. 

Another important kinematical variable is the momentum transfer squared, $t=(p_2-p_4)^2$, where $p_2$ and $p_4$ are the four-momenta of the target and of the scattered nucleon. At threshold, one can find that the variable $t$ has only a fixed value, $t=-mm_V$, where $m$ is the nucleon mass and $m_V$ is the $V-$meson mass. So, for $J/\psi$-production this momentum is large: $t\simeq$ -3 GeV$^2$, therefore all the propagators corresponding to the light mesons are of comparable magnitude ($t-m^2_{\pi}$ $\simeq$ $t-m^2_{\eta}$ $\simeq$ 
$t-m^2_ {\sigma}$ $\simeq$ $t-m^2_{\rho}$). In such situation it is not possible to justify the dominance of a particular exchange mechanism, so we have to conclude that threshold heavy $V$-meson production in NN-collisions is determined by the exchange of the coherent sum of many different mesons, with different masses. 

But what about the quantum numbers, ${\cal J}^P$, of these exchanges 
(${\cal J}$ is the spin and $P$ is the parity of the corresponding meson)? 

We can use the parametrizations  (\ref{eq:mpp}) and (\ref{eq:mnp}), which are exact and model independent results, with definite selection rules, from the point of view of $s$-channel for $N+N\to N+N+V$ to understand the $t-$channel ${\cal J}^P$-picture.
Using the Fierz-transformations (in two-component form), let us rewrite the general parametrization of the matrix element, see Eqs. (\ref{eq:mpp}) and (\ref{eq:mnp}), as a $t$-channel parametrization:
\begin{eqnarray}
&(\tilde{\chi}_2~\sigma_y \chi_1) (\chi^{\dagger}_4 \vec{\sigma}
\cdot\vec U^* \times\hat{\vec k }\sigma_y{\chi}^{\dagger}_3 )=&
\displaystyle\frac{1}{2}({\chi}^{\dagger}_3 \chi_1)~(\chi^{\dagger}_4\vec{\sigma}
\cdot\vec  U^* \times\hat{\vec k } {\chi}_2)+\displaystyle\frac{1}{2}({\chi}^{\dagger}_3\vec{\sigma}
\cdot\vec  U^* \times\hat{\vec k }\chi_1)~(\chi^{\dagger}_4{\chi}_2)+\label{eq:fierz} \\
&&\displaystyle\frac{1}{2}\left [( {\chi}^{\dagger}_3\vec{\sigma}\cdot\vec  U\chi_1)~(
\chi^{\dagger}_4\vec{\sigma}\cdot\hat{\vec k }{\chi}_2)-( {\chi}^{\dagger}_3\vec{\sigma}\cdot\hat{\vec k }\chi_1)~(
\chi^{\dagger}_4\vec{\sigma}\cdot\vec U^*{\chi}_2)\right ].\nonumber 
\end{eqnarray}
Each term in Eq. (\ref{eq:fierz}) has a precise dynamical interpretation, as it corresponds to $t-$channel meson exchange (Fig. \ref{fig:fig1}) with a definite spin and parity, ${\cal J}^P$. The first two terms describe a scalar exchange, where the spin structure 
$\vec{\sigma}\cdot\vec  U^* \times\hat{\vec k }$ (in one vertex) corresponds to the matrix element of the process $\sigma^*+N\to V^0+N$ ($\sigma^*$ is a virtual scalar meson) at its threshold. The other vertex, corresponding to the $\sigma NN$-interaction has a structure of the type $\chi^{\dagger} 
{\cal I}\chi$. The last two terms describe the exchange by neutral pseudoscalar mesons ($\pi^0$ or  $\eta$) with the  $\vec{\sigma}\cdot\hat{\vec U }$-spin structure of the matrix element for the subprocess ${\pi^0}^*(\eta^*)+N\to V^0+N$ at threshold, and with the $\vec{\sigma}\cdot\hat{\vec k }$ structure for the vertex $NN\pi(\eta)$. The same considerations hold for the $f_{10}$-partial amplitude. 

Therefore we can conclude that $t$-channel exchanges with ${\cal J}^P=0^+$ (scalar mesons) and ${\cal J}^P=0^-$ (pseudoscalar meson) can be considered as the most probable mechanisms to describe the threshold dynamics of $J/\psi$-production in NN-collisions. Let us consider these mechanisms in detail.
\subsection{$\sigma$-exchange}

The matrix element ${\cal M}_{\sigma}$, corresponding to 
the two diagrams of Fig. \ref{fig:fig1}, can be written as:
$$
{\cal M}_{\sigma}= {\cal M}_{1\sigma}+{\cal M}_{2\sigma},$$
with the following expression for the matrix element ${\cal M}_{1\sigma}$,  corresponding to Fig. \ref{fig:fig1}a:
\begin{equation}
{\cal M}_{1\sigma}= -\displaystyle\frac{g_{\sigma NN}}{t-m^2_{\sigma}}{\cal N}
({\chi}^{\dagger}_3 {\cal I}\chi_1)\left [\chi^{\dagger}_4(ih_{1\sigma}
\hat{\vec k }\cdot\vec  U^*+h_{2\sigma}\vec{\sigma}\cdot\hat{\vec k }\times
\vec  U^*) \chi_2\right ],
\label{eq:ms1}
\end{equation}
where $g_{\sigma NN}$ is the $\sigma NN$ coupling constant, ${\cal N}=2m(E+m)=m(m_V+4m)$ is a normalization factor, which arises from the transformation of the invariant 
matrix element for the considered process (in terms of the four-component spinors for the initial and final nucleons) to the two-component form, which is better adapted to the description of threshold spin structure in CMS of the considered process. We used the following formula for the threshold energy of the initial nucleons, $E=\displaystyle\frac{W}{2}=m+\displaystyle\frac{m_V}{2}$. The complex amplitudes $h_{1\sigma}$ and $h_{2\sigma}$ describe two possible threshold partial transitions in $\sigma^*+N\to V+N$: 
$\ell_{\sigma}=1\to {\cal J}^P=1/2^-$ and $3/2^-$, where  $\ell_{\sigma}$ is the orbital momentum of the initial $\sigma N$-system. One can find that: 
$$h_{1\sigma}=h_{1/2}+2h_{3/2},~h_{2\sigma}=h_{1/2}-h_{3/2},$$
where $h_{1/2}$ and $h_{3/2}$ are the partial amplitudes corresponding to the two possible values of the total angular momentum in $\sigma^*+N\to V+N $.

Comparing the spin structure of the matrix elements for the processes 
$n+p\to n+p+V^0$, Eq. (\ref{eq:mnp}) and $\sigma+N\to N+V^0$, Eq. (\ref{eq:ms1}), one can see that only the amplitude $h_{2\sigma}$ has to be kept, because it generates transversally polarized $V-$mesons. First of all, this means that the cross section of the processes $n+p\to n+p+V^0$
and $\sigma +N\to N+V$ are determined by different contributions to the amplitudes $h_{1\sigma}$ and $h_{2\sigma}$. This follows from the following formula for the corresponding cross section: 
$$\displaystyle\frac{d\sigma}{d\Omega}(\sigma N)\simeq |h_{1\sigma}|^2+2|h_{2\sigma}|^2, 
~\displaystyle\frac{d\sigma}{d\Omega}(np\to npV)\simeq |h_{2\sigma}|^2.$$
Therefore these cross sections must be independent. Moreover, both amplitudes $f_{01}$ and $f_{10}$, for $n+p\to n+p+V^0$ in case of $\sigma$-exchange, being proportional to $h_{2\sigma}$, have to satisfy simple relations. In such case definite numerical values can be derived for the polarization observables in $V-$meson production for $np-$collisions. To prove this, let us transform the matrix element (\ref{eq:ms1}) into the "standard"  parameterization of Eq. (\ref{eq:mnp}), in terms of definite quantum numbers of the $s$-channel:
\begin{eqnarray}
&-(\chi^{\dagger}_3\chi_1)~(\chi^{\dagger}_4 \vec{\sigma}
\cdot\vec A{\chi}_2)=&
\displaystyle\frac{1}{2}\left [-(\tilde{\chi}_2~\sigma_y ~\vec{\sigma}
\cdot\vec A\chi_1)~(\chi^{\dagger}_4
\sigma_y \tilde{\chi}^{\dagger}_3 )+\right .\nonumber \\
&& (\tilde{\chi}_2~\sigma_y\chi_1)~(\chi^{\dagger}_4\vec{\sigma}
\cdot\vec A \sigma_y \tilde{\chi}^{\dagger}_3 )+\label{eq:mppf} \\&&
\left .+i\epsilon_{ikl}A_i(\tilde{\chi}_2~\sigma_y ~\sigma_{\ell} \chi_1)~(\chi^{\dagger}_4 \sigma_k\sigma_y\tilde{\chi}^{\dagger}_3 )\right ],\nonumber 
\end{eqnarray}
where $\vec A= \hat{\vec k }\times\vec U^*$. From Eq. (\ref{eq:mppf}) one can see that ${\cal M}_{1\sigma}$ contains not only the structures which are  allowed by symmetry selection rules, but also a contribution which corresponds to a triplet-triplet transition in the $np$-system (last term in Eq. (\ref{eq:mppf})). Such transition is forbidden by the generalized Pauli principle, following from the isotopic invariance of the strong interaction and should not appear in the total matrix element ${\cal M}_\sigma$. 

Let us consider in a similar way the matrix element ${\cal M}_{2\sigma}$:
\begin{equation}
{\cal M}_{2\sigma}= -\displaystyle\frac{g_{\sigma NN}}{t-m^2_{\sigma}}{\cal N}
\left [ {\chi}^{\dagger}_3 \left(ih_{1\sigma}
\hat{\vec k }\cdot\vec  U^*+h_{2\sigma}\vec{\sigma}\cdot\hat{\vec k }\times
\vec  U^*\right ) \chi_1\right ] \left (\chi^{\dagger}_4{\cal I}\chi_2\right ),
\label{eq:ms2}
\end{equation}
where we applied the following relations: $g_{\sigma pp}=g_{\sigma nn}=g_{\sigma NN},$ $h_{i\sigma}(\sigma n\to nV^0)=
h_{i\sigma}(\sigma p\to pV^0),~i=1,2$, which follow from the isotopic invariance of the strong interaction, in case of isoscalar vector meson. Note that here we use the same propagator as in Eq. (\ref{eq:ms1}). this is correct in threshold conditions, because any different propagator will generate higher waves in the initial and final states. Summing the two contributions in the matrix element, ${\cal M}_{\sigma}$, the $wrong$ term corresponding to the triplet-triplet transition in $n+p\to n+p+V^0$ is exactly cancelled:
$$
{\cal M}_{1\sigma}+{\cal M}_{2\sigma}=(\tilde{\chi}_2~\sigma_y ~\vec{\sigma}\cdot\vec A\chi_1)~(\chi^{\dagger}_4\sigma_y \tilde{\chi}^{\dagger}_3 )+(\tilde{\chi}_2~\sigma_y\chi_1)~(\chi^{\dagger}_4\vec{\sigma}
\cdot\vec A\sigma_y \tilde{\chi}^{\dagger}_3 ),
$$
Therefore, for $\sigma-$exchange, one finds:
\begin{equation}
f_{10}^{(\sigma)}=-f_{01}^{(\sigma)}=-\displaystyle\frac{g_{\sigma NN}}{t-m^2_{\sigma}}{\cal N}h_{2\sigma}.
\label{eq:ffs}
\end{equation}
Independently on the numerical values of the coupling constant $g_{\sigma NN}$
and of the partial amplitude $h_{2\sigma}$ for the process $\sigma +N\to N+V^0$, the polarization observables for the process $n+p\to n+p+V^0$ and the ratio ${\cal R}$ (see Eq. (\ref{eq:eqr})),  take the following values (in framework of $\sigma$-exchange):
\begin{equation}
2{\cal A}_1^{(\sigma)} (np)=P_1^{(\sigma)}(np)=-1, ~\mbox{and~}  ~{\cal R}^{(\sigma)}=1.
\label{eq:aos}
\end{equation}
Note that introducing phenomenological form factors in the expression for ${\cal M}_{\sigma}$ affects the absolute value of the cross section, but can not change the relation $f_{10}^{(\sigma)}=-f_{01}^{(\sigma)}$ and therefore the results (\ref{eq:aos}).

\subsection{$\eta$-exchange}

\begin{figure}
\mbox{\epsfysize=15.cm\leavevmode \epsffile{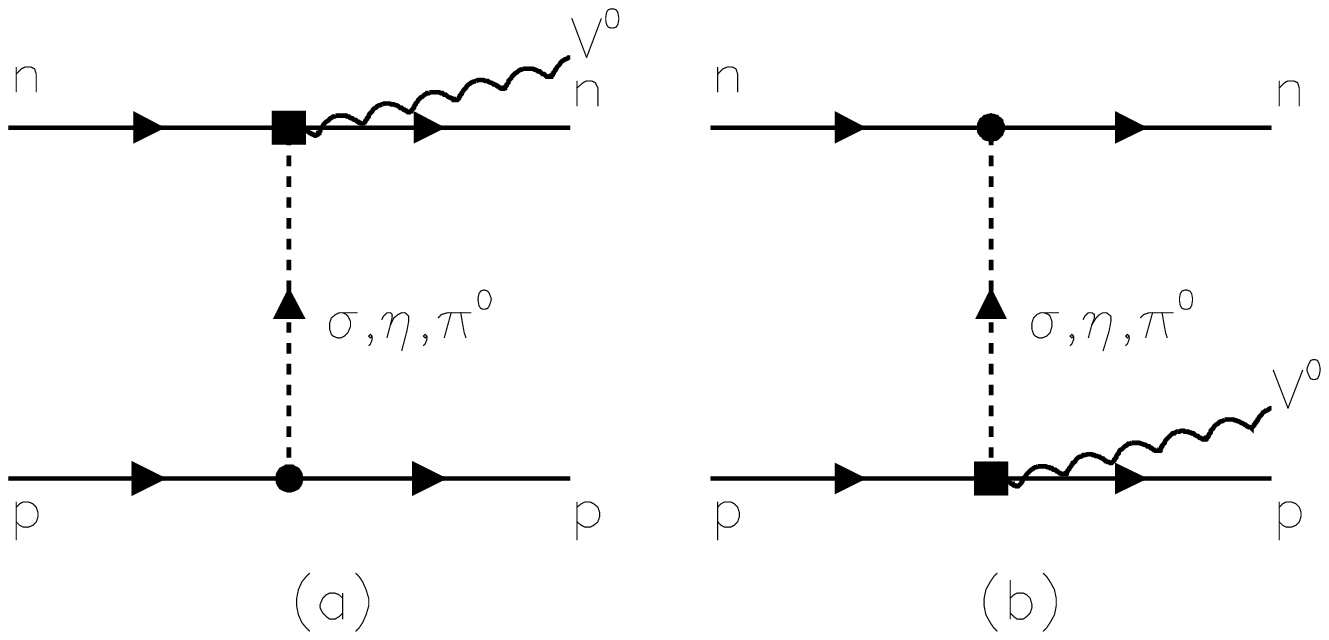}}
\vspace*{.2 truecm}
\caption{Feynman diagrams for the process 
$n+p\to n+p+V^0$, describing $t-$channel exchanges by neutral $\sigma$, $\eta$
and $\pi^0$-mesons.
}
\label{fig:fig1}
\end{figure}

Similarly to $\sigma$-exchange, the $\eta$-exchange is characterized by two diagrams (Fig. \ref{fig:fig1}) and the matrix element is the sum of two matrix elements, corresponding to Fig. \ref{fig:fig1}a  and Fig. \ref{fig:fig1}b:
$$
{\cal M}_{\eta}= {\cal M}_{1\eta}+{\cal M}_{2\eta},$$
where the matrix element ${\cal M}_{1\eta}$ can be written as:
\begin{equation}
{\cal M}_{1\eta}=-\displaystyle\frac{g_{\eta NN}}{t-m^2_{\eta}}
\left( \displaystyle\frac{k}{E+m}\right ) {\cal N}
\left [\chi^{\dagger}_3 (h_{1\eta}\vec{\sigma}\cdot\vec  U^*+h_{2\eta}
\vec{\sigma}\cdot\hat{\vec k }\cdot \vec  U^*) \chi_1\right ](\chi^{\dagger}_4\vec{\sigma}\cdot\hat{\vec k }\chi_2),
\label{eq:me1}
\end{equation}
where $h_{1\eta}$ and $h_{2\eta}$ are two independent partial amplitudes, which describe the threshold spin structure for the subprocess $\eta^*+N\to V+N^0$. These amplitudes correspond to to two allowed threshold partial transitions (in $\eta^*+N\to V+N^0$): $\ell_i=0\to {\cal J}^P=1/2^-$ and $\ell_i=2\to {\cal J}^P=3/2^-$.
The factor  $\displaystyle\frac{k}{E+m}=\sqrt{\displaystyle\frac{m_V}{m_V+4m}}$ arises from the transformation from the relativistic expression of the $\eta NN$-vertex, $\overline{u}(p_2) \gamma_5 u(p_1)$, to the two-component form in the CMS of $n+p\to n+p+V$-reaction, $\vec\sigma\cdot\hat{\vec k }$. 

Again, one can prove that only the sum ${\cal M}_{1\eta}+{\cal M}_{2\eta}$ generates the correct spin structure for threshold matrix element ${\cal M}_{\eta}$ (again taking into account the isotopic invariance for both vertexes of the considered  diagrams: $\eta NN$ and $\eta  +N\to N+V^0$, $N=p$ or $n$):
\begin{eqnarray}
&{\cal M}_{1\eta}+{\cal M}_{2\eta}=\displaystyle\frac{g_{\eta NN}}{t-m^2_{\eta}}
{\cal N}\sqrt{\displaystyle\frac{k}{E+m}}h_{1\eta}&
\left [(\tilde{\chi}_2~\sigma_y ~\vec{\sigma}\cdot\vec A\chi_1)~(\chi^{\dagger}_4\sigma_y \tilde{\chi}^{\dagger}_3 )+\right .\nonumber \\
&&\left .(\tilde{\chi}_2~\sigma_y\chi_1)~(\chi^{\dagger}_4\vec{\sigma}
\cdot\vec A\sigma_y \tilde{\chi}^{\dagger}_3 )\right  ],
\label{eq:mppe}
\end{eqnarray}
with the following relation for the partial amplitudes for $n+p\to n+p+\eta$:
\begin{equation}
f_{10}^{(\eta)}=f_{01}^{(\eta)}
\label{eq:ffe}
\end{equation}
and definite numerical predictions for the polarization phenomena and for the ratio ${\cal R}$:
\begin{equation}
{\cal A}^{(\eta)}_{1np}=\displaystyle\frac{1}{2},~{\cal P}^{(\eta)}_{1np}=1~\mbox{and~}  ~{\cal R}^{(\eta)}=1.
\label{eq:pos}
\end{equation}
So, only the coefficient ${\cal P}^{(\eta)}_{1np}$ can discriminate between $\eta$- and $\sigma$- exchanges.

\subsection{$\pi$-exchange}
Due to the isotopic invariance of the strong interaction, it is necessary to consider four Feynman diagrams, corresponding to the exchange of neutral and charged pions in $n+p\to n+p+V^0$ (see fig. 2). Taking into account the isotopic relations between different coupling constants in $NN\pi$-vertexes and different amplitudes for the processes $\pi+N\to N+V^0$, one can find the following expressions for the amplitudes $f_{10}^{(\pi)}$ and $f_{01}^{(\pi)}$:
\begin{equation}
f_{10}^{(\pi)}=-\displaystyle\frac{g_{\pi NN}}{t-m^2_{\pi}}h_{1\pi} {\cal N}
\sqrt{\displaystyle\frac{m_V}{4m+m_V}},~f_{01}^{(\pi)}=-3f_{10}^{(\pi)},
\label{eq:pop}
\end{equation}
and the single amplitude $f_{10}^{(\pi)}$ for the process $p+p\to p+p+V^0$ is equal to 2$f_{10}^{(\pi)}(np\to npV^0)$. Note that the spin structure for the processes $\pi+N\to N+V$ and $\eta+N\to N+V$ has to be simpler.

Independently from the concrete model for the amplitude $h_{1\pi}$ of the process $\pi+N\to N+V$, the relations (\ref{eq:pop}) allow to predict definite values for the polarization observables  and for the ratio ${\cal R}$:
\begin{equation}
{\cal A}^{(\pi)}_{1np}=-\displaystyle\frac{9}{10},~{\cal P}^{(\pi)}_{1np}=-\displaystyle\frac{3}{5} ~\mbox{and~}  ~{\cal R}^{(\pi)}=5.
\label{eq:popr}
\end{equation}
which are very different from the previous cases of pure $\sigma$- or  $\eta$-exchanges.
\subsection{"Realistic model": $\pi+\eta$-exchange}
Based on the above mentioned results for the different $t-$channel exchanges, we can build a more realistic model, combining the contributions of different mesons.
As an example, let us consider the case of $\pi+\eta$-exchange, with the following expressions for the allowed threshold partial amplitudes $f_{10}$ and $f_{01}$ of the process $n+p\to n+p+V^0$:
\begin{equation}
f_{10}=-{\cal A}_{\pi}(1-r),~f_{01}={\cal A}_{\pi}(3+r),
\label{eq:popa}
\end{equation}
where the ratio $r=\displaystyle\frac{g_{\eta NN}}{g_{\pi NN}}\displaystyle\frac{h_{1\eta}}{h_{1\pi}}\left (\displaystyle\frac {t-m^2_{\pi}}{t-m^2_{\eta}}\right )$ characterizes the relative role of $\eta$- and $\pi$-exchanges in  $n+p\to n+p+V^0$ and 
${\cal A}_{\pi}=\displaystyle\frac{g_{\pi NN}}{t-m^2_{\pi}}\sqrt{\displaystyle\frac{m_V}{m_V+4m}}h_{1\pi}$.
Therefore, we can find the following results for the polarization observables in $n+p\to n+p+V^0$ and for 
the ratio ${\cal R}$ of the total cross section for $n+p$- and $p+p$-collisions:  \begin{equation}
{\cal A}_1=-\displaystyle\frac{9+6{\cal R}e ~r +|r|^2}{2(5+2{\cal R}e ~r +|r|^2)}, 
~{\cal P}_1=-\displaystyle\frac{3-2{\cal R}e ~r -|r|^2}{5+2{\cal R}e ~r +|r|^2}, ~{\cal R}=\displaystyle\frac{5+2{\cal R}e ~r +|r|^2}{|1-r|^2}, 
\label{eq:pos1}
\end{equation}
This primarily means that, in framework of the considered model, two independent parameters ${\cal R}e~r$ and $|r|^2$ enter in the definition of three observables. Therefore a simultaneous measurement of ${\cal P}_1$ and 
${\cal A}_1$ can determine uniquely ${\cal R}e ~r$ and $|r|^2$ (with the evident condition $|{\cal R}e~r|<|r|$):
\begin{equation}
{\cal R}e ~r =-1- 2\displaystyle\frac{1+2{\cal A}_1}{1-{\cal P}_1},~|r|^2=-3+
4\displaystyle\frac{3+2{\cal A}_1}{1-{\cal P}_1}
\label{eq:pos2}
\end{equation}

The situation is simplified if the ratio $r$ is a real parameter. This is the case in framework of the effective Lagrangian approach for the processes
$\eta(\pi)+N\to N+V^0$, near threshold, where the corresponding pole Feynmann diagrams originate the real amplitudes $h_{1\eta}$ and $h_{1\pi}$. It is also the case for $s-$channel $N^*$-contribution, which is common to $\eta +N$ and $\pi+N$- interactions. For $J/\psi$-production the first case seems as the most probable. 

For a real value of $r$, the following quadratic relation between polarization observables ${\cal P}_1$ and ${\cal A}_1$ holds:
\begin{equation}
4 A_1+4{\cal A}_1^2+{\cal P}_1^2=0.
\label{eq:pos3}
\end{equation}

After measuring of the ratio ${\cal R}$ (of total cross sections for $np-$ and $pp-$ interactions) it will be possible to find two different values for $r$:
$$r_{\pm}=\displaystyle\frac{{\cal R}+1}{{\cal R}-1}\pm 2\displaystyle\frac{\sqrt{2{\cal R}-1}}{{\cal R}-1}.
$$

Knowing $r$, it is straightforward to predict any polarization observable for the process $n+p\to n+p+V^0$. The behaviors of ${\cal R}$, ${\cal A}_1$ and ${\cal P}_1$ as a function of  $r$, (when $r$ is real) are shown in Figs. 3-5.
One can see a strong dependence of the ratio ${\cal R}$ from $r$ in the region $-1\le r\le 3$, where ${\cal R}> 1$, i.e. with strong isotopic effects. Only for $r<-1$ we have ${\cal R}<1$, with a weak dependence on the parameter $r$. The coefficients ${\cal A}_1$ and ${\cal P}_1$ show particular sensitivity to $r$.

\subsection{Attempts to estimate $r$}
The previous analysis is based on the most general properties of threshold vector meson production in $NN$-collisions and $t$-channel exchanges. In the last case, we used only properties related to the quantum numbers of the corresponding $t$-channel mesons: spin, parity and isotopic spin. All previous results are valid for any isoscalar vector meson production, $\omega$, $\phi$ or  $J/\psi$. The properties, which are specific to a definite $V^0$-meson reaction, appear first of all in the kinematics, in particular in the different $V^0$-meson masses, and in the values of the partial amplitudes for the binary subprocess,
$M^*+N\to N+V^0$, where $M^*=\sigma,\eta,\pi$ is a virtual meson (with space-like four-momentum). So, for the considered $(\eta+\pi)$-model, all dynamical information, which is necessary for the calculations of such observables as ${\cal R}$, ${\cal A}_1$ and ${\cal P}_1$, is contained in one complex parameter $r$, which characterizes the ratio of the threshold amplitudes for $\eta(\pi)+N\to N+V^0$-processes. Therefore the identity of $V^0$-production is also contained in the ratio $r$. 

Let us estimate this ratio in case of $J/\psi$- production. For this aim, we refer to the existing experimental information about the different hadronic decays of the $J/\psi$- meson. For example, the following branching ratios \cite{PdG}:
$$BR(J/\psi\to \rho^0\pi^0)=(4.2\pm 0.5)\cdot 10^{-3},$$
$$BR(J/\psi\to \omega\eta)=(1.58\pm 0.16)\cdot 10^{-3},$$
allow to determine $r$ in the framework of a simple vector exchange model for the process $\eta(\pi)+N\to N+N+J/\psi$ (see Fig. 6). The nice property of this model is that all the coupling constants are known. The $\rho-$exchange mechanism for $J/\psi$-production in $\pi N$-collisions has been considered earlier \cite{Br97}. The corresponding matrix element can be written in the following form:
\begin{equation}
{\cal M}=
\displaystyle\frac{g_{PVV'}}{t-m^2_{V'}}\displaystyle\frac{\epsilon_{\mu\nu\rho}
U_{\mu}k_{\nu}q_{\rho}J_{\sigma}}{m_{J/\psi}}, 
\label{eq:mjp}
\end{equation}
\begin{eqnarray}
&J_{\sigma}&=\overline{u}(p_2)\left [\gamma_{\sigma}F_1-\displaystyle\frac{\sigma_{\sigma\nu}(k-q)_{\nu}}{2m}F_2\right ]u(p_1)\nonumber \\
&&=\overline{u}(p_2)\left [\gamma_{\sigma}(F_1+F_2)-\displaystyle\frac{p_{1\sigma}+p_{2\sigma}}
{2m}F_2\right ]u(p_1),
\label{eq:ff}
\end{eqnarray}
where $F_1$ and $F_2$ are the Dirac and Pauli form factors of the $V'NN$-vertex of the considered diagram. At the reaction threshold the matrix element, Eqs. (\ref{eq:mjp}) and (\ref{eq:ff}), 
can be simplified as:
\begin{equation}
{\cal M}\simeq \displaystyle\frac{g_{PVV'}}{t-m^2_{V'}}(F_1+F_2)\chi^{\dagger}_2\left (
\vec\sigma\cdot \vec U^*-\vec\sigma\cdot\hat{\vec k }\hat{\vec k }\cdot\vec U^*\right )\chi_1.
\label{eq:mjp1}
\end{equation}
Taking into account VDM-predictions for the $(F_1+F_2)$ term, which in case of $\omega NN(\rho NN)$-vertex, is proportional to the isoscalar, $\mu_p+\mu_n$, (isovector, $\mu_p-\mu_n$,) magnetic moment of the nucleon,  one can find:
$$r=\displaystyle\frac{g_{\eta NN}}{g_{\pi NN}}\displaystyle\frac{(\mu_p+\mu_n)}
{(\mu_p-\mu_n)}\displaystyle\frac{g(J/\psi\to \eta\omega)}{g(J/\psi\to \pi^0\rho^0)}\simeq \displaystyle\frac{g_{\eta NN}}{g_{\pi NN}}
\displaystyle\frac{0.88}{4.8}\sqrt{\displaystyle\frac{\Gamma(J/\psi\to \eta\omega)}{\Gamma (J/\psi\to \pi^0\rho^0)}}\simeq \displaystyle\frac{1}{7}
\left |\displaystyle\frac{g_{\eta NN}}{g_{\pi NN}}\right |,$$
where $g$ ($\Gamma$) is the corresponding coupling constant (decay width).
All existing analysis of threshold $\eta-$meson photoproduction on protons 
\cite{Ch01} indicate that $g_{\eta NN}\ll g_{\pi NN}$. Therefore, in the framework of such model, we have $|r|\ll 1$. From these considerations we can not determine the sign of the ratio $r$, but the small value of $|r|$ indicates that the $np$-cross section can be essentially larger than $pp$-cross section, in case of 
$J/\psi$ production.

However conclusions could be different if one takes another set of experimental data about $J/\psi$-decays \cite{PdG}:
$$BR(J/\psi\to p\overline{p}\eta)=(2.09\pm 0.18)\cdot 10^{-3},$$
\begin{equation}
BR(J/\psi\to p\overline{n}\pi^-)=(2.00\pm 0.10)\cdot 10^{-3},
\label{eq:br2}
\end{equation}
Evidently both these decays can be considered as crossed-channels of the processes $\eta(\pi)+N\to N+J/\psi $. Generally, each decay $J/\psi\to N+\overline {N}+P$ ($P$ is the pseudoscalar meson, $P=\pi$ or $\eta$) is characterized by a complicated spin structure - with six independent scalar (and complex) amplitudes, which are functions of two independent kinematical variables. It is not possible to restore the full spin structure, from the knowledge of the branching ratio, alone (with unpolarized particles). Moreover,  there is the delicate problem of the extrapolation from the decay region of the kinematical variables (of the process $J/\psi\to N+\overline {N}+P$) to the scattering region (of the process $P+N\to N+J/\psi$). To overcome this problem, let us consider the oversimplified assumption that the two types of processes $J/\psi\to N+\overline {N}+P$ and $P+N\to N+J/\psi$ are driven by an effective contact four-particle interaction, with a single coupling constant. The exact spin structure of this interaction is not important for our considerations. In such approximation the ratio $r$ can be estimated from the following formula:
$$r=\displaystyle\frac{g_{\eta NN}}
{g_{\pi NN}}\displaystyle\frac{(t-m_{\pi}^2)}{(t-m_{\eta}^2)}
\left [\displaystyle\frac{BR(J/\psi\to N\overline{N}\eta)}
{BR(J/\psi\to N\overline{N}\pi^-)}\right ]^{1/2}\simeq \displaystyle\frac{g_{\eta NN}}{g_{\pi NN}}< 1 $$
Again the sign of $r$ can not be determined by such considerations, but, again the value of $r$ which has been derived is in the range where the ratio ${\cal R}$ is very sensitive to the value of $r$. Note that the dependence of the polarization observables ${\cal A}_1$ and ${\cal P}_1$  is also quite large, in this region of $r$.

In the previous considerations, the effects of the final state interaction in the produced $NNJ/\psi$-three particle system were not taken into account. In the near threshold region, the $NN-$interaction is well known, in terms of the corresponding scattering energies and effective radius. It is not the case for the $J/\psi N$-interaction. Note in this connection, that the total 
$J/\psi N$-cross section is not presently well known \cite{Mu99}. For example, photoproduction data give values of 3-4 mb \cite{Hu98}, while the analysis of charmonium absorption on nucleons (at relatively high momentum) in $p+A$- and $A+A$-reactions suggests larger values, 6-7 mb \cite{Wa98,Kh97}. These larger values can be explained in the framework of effective Lagrangian approaches \cite{Ma98,Ha99,Si00}. Different methods have been suggested for a direct measurement of this quantity, through the processes $\pi+d\to J/\psi + p+p$ \cite{Br97}, $\overline{p}+d\to J/\psi + n$ \cite{Ca00}, and 
charmed meson production in $\overline{p}+A$-collisions \cite{Go84}.

Note, that in the general case, it is necessary to define two different 
$J/\psi$-cross sections corresponding to transversal and longitudinal $J/\psi$-polarization. For example, the data about $\gamma+N\to J/\psi+N$ are sensitive to $\sigma_T(J/\psi N)$-cross section, with transversal 
$J/\psi$-polarization, if the VDM hypothesis is correct. Another possible method is to determine the average 
$\sigma_{Av}(J/\psi N)$-cross section, obtained by averaging over the $J/\psi$-polarization. In the case of interest here, the near-threshold $J/\psi$-production in nucleon-nucleon collisions, $N+N\to N+N+J/\psi$, the possible effects of the $J/\psi N$-interaction are given by the $\sigma_T(J/\psi N)$-cross section alone, because we showed that the kinematical conditions are such that the $J/\psi$ is produced with transversal polarization, only. To avoid double counting in the calculation of the $ J/\psi N$ final interaction, one has to take into account only the $ J/\psi N$ interaction with a nucleon spectator, produced in the vertex $\pi NN$.

Let us compare the cross sections for the $\phi$ and $J/\psi $- production in $pp$-collisions - in the framework of the same approach, namely for $\pi$-exchange in $N+N\to N+N+V^0$ and $\rho$-exchange for the subprocess $\pi+N\to N+V^0$, with $V^0=\phi$ or $J/\psi $. For the same value of $Q=\sqrt{s}-2m-m_V$, we can write (in the near-threshold region):
$$R(J/\psi,\phi)=\displaystyle\frac{\sigma(pp\to ppJ/\psi)}{\sigma(pp\to pp\phi)}\simeq
\displaystyle\frac{g^2(J/\psi\to\pi\rho)}{g^2(\phi\to\pi\rho)}\left (\displaystyle\frac{t_{\phi}-m_\pi^2}{t_{J/\psi}-m_\pi^2}\right )^2
\left [\displaystyle\frac{F(t_{J/\psi})}{F(t_{\phi})}\right ]^2,$$
where $g(V\to \pi\rho)$ is the coupling constant for the decay $V\to \pi\rho$, $t_V=-mm_V$ is the threshold value of the momentum transfer squared, $F(t)$ is a phenomenological form factor for the vertex $\pi^*\rho^*V$, with virtual $\pi$ and $\rho$. Using the existing experimental data about the decays 
$J/\psi\to \pi+\rho$ and $\phi\to \pi+\rho $ , one can find $g^2(J/\psi\to\pi\rho)/g^2(\phi\to\pi\rho)\simeq 10^{-4}$, so that $R(J/\psi,\phi)\simeq 10^{-5}\left [\displaystyle\frac{F(t_{J/\psi})}{F(t_{\phi})}\right ]^2$. Taking into account  that $\sigma(pp\to pp\phi)\simeq $ 300 nb at $p_L=3.67$ GeV \cite{Ba01}, one can find  that
$\sigma(pp\to ppJ/\psi)\simeq $  0.03 nb $\left [\displaystyle\frac{F(t_{J/\psi})}{F(t_{\phi})}\right ]^2$. This value is too small, when compared with the existing experimental value for the lowest $\sqrt{s}=6.7$ GeV, namely \cite{Ba78} $\sigma_{exp}(pp\to ppJ/\psi)=0.3\pm 0.09$ nb.

Note that the $\rho$-exchange model for $\sigma(\pi N\to J/\psi)$ gives a cross section one order of magnitude smaller in comparison with other possible theoretical approaches \cite{Ba78,Bo75,Ko79}. It is one possibility to explain the value of $\sigma_{exp}(pp\to ppJ/\psi)$. Another possibility is to take
$\left [\displaystyle\frac{F(t_{J/\psi})}{F(t{\phi})}\right ]^2\simeq 10$, which can be plausible, because the $J/\psi=c\overline{c}$-system must have a smaller size in comparison with $\phi=s\overline{s}$. This can be realized by the following form factor:
$$F_V(t)= \displaystyle\frac{1}{1-\displaystyle\frac{t}{\Lambda_V^2}}$$
with ${\Lambda_V}\simeq m_V$.

\section{Conclusions}
Let us summarize the main results concerning the theoretical analysis of $J/\psi$-production in nucleon-nucleon collisions in threshold regime.
\begin{itemize}
\item We established the spin structure of the threshold matrix element in terms of a limited number of partial transitions, corresponding to $S$-wave production of final particles in the process $N+N\to N+N+J/\psi$.
\item We proved the essential role of isotopic effects for $J/\psi$-production in $pp$- and $np$-collisions. The two reactions present very different characteristics concerning:
\begin{itemize}
\item the number of independent partial transitions,
\item the spin structure of the threshold matrix elements,
\item the value of the absolute cross sections,
\item the polarization phenomena.
\end{itemize}
Note that all these differences are generated by a common mechanism: the origin of the essential difference has to be found in the different role of the Pauli principle for $pp$- and $pn$-collisions in the near threshold region.

\item This model-independent analysis shows the universality of theoretical considerations of threshold production of different vector mesons in nucleon-nucleon collisions, starting from $light$ $\rho$, $\omega$, $\phi$ to charmed mesons.

\item Only one polarization observable, the  $J/\psi$-polarization, is identical for  $pp$- and $pn$-collisions: the  $J/\psi$-meson is transversally polarized -even in collisions of unpolarized nucleons. The experimental determination of the ratio of the total cross sections for $np-$ and $pp-$collisions is important for the identification of the reaction mechanism. Polarization phenomena, which are trivial for threshold $pp$-collisions, will be very useful for $np$-collisions. The polarization transfer coefficients show the largest sensitivity to the nature (quantum numbers) of $t-$channel exchanges.

\item The existing experimental data  on the specific decays of $J/\psi$, such as 
$J/\psi\to \rho\pi$, $J/\psi\to \omega\eta$, $J/\psi\to N\overline{N}\pi$, $J/\psi\to N\overline{N}\eta$, constrain a possible model for the threshold $J/\psi$-production in nucleon-nucleon collisions. These data show, in a transparent way, the link between the possible reaction mechanisms on one side, and the physics of $J/\psi$-decays in usual hadrons ($N$, $\overline{N}$, $\pi$, $\rho$...), on the other side.

\end{itemize}

\begin{figure}
\mbox{\epsfysize=15.cm\leavevmode \epsffile{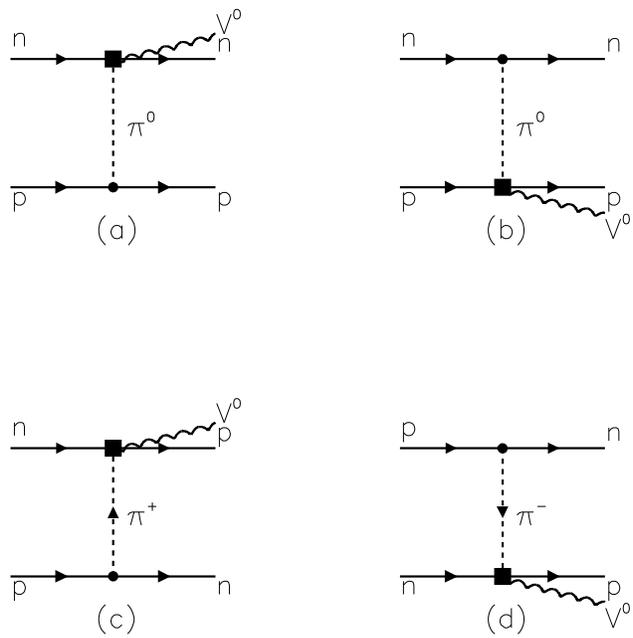}}
\vspace*{.2 truecm}
\caption{Pion exchange for the process $n+p\to n+p+V^0$.}
\label{fig:fig2}
\end{figure}

\begin{figure}
\mbox{\epsfysize=15.cm\leavevmode \epsffile{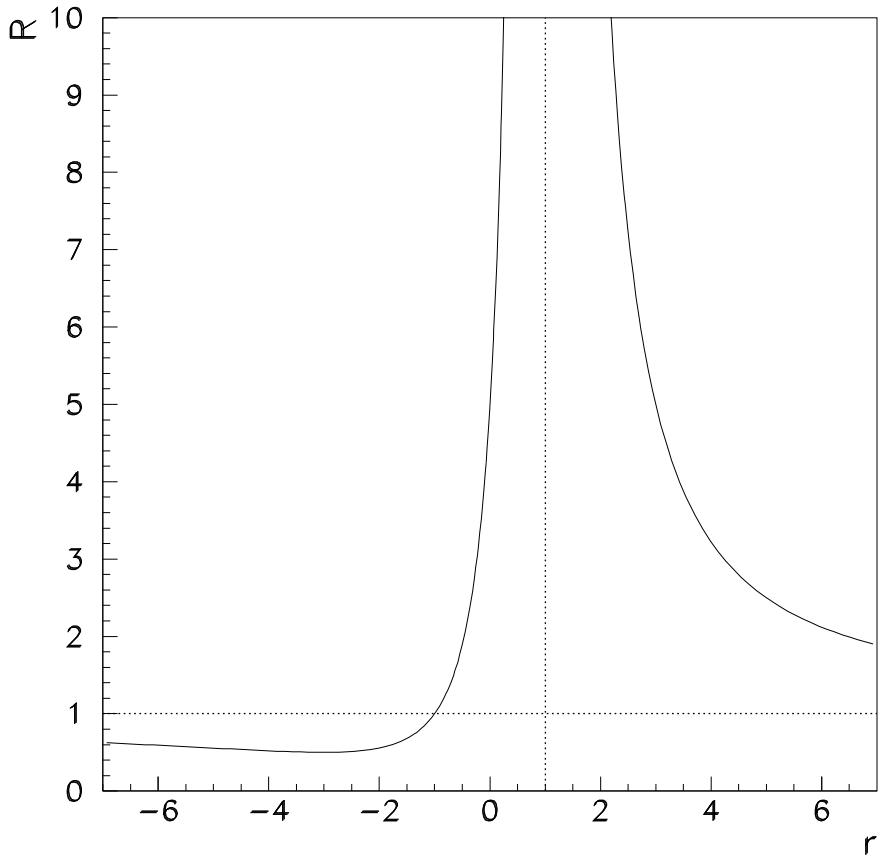}}
\vspace*{.2 truecm}
\caption{Dependence of ${\cal R}$ on $r$}.
\label{fig:fig3}
\end{figure}
\begin{figure}
\mbox{\epsfysize=15.cm\leavevmode \epsffile{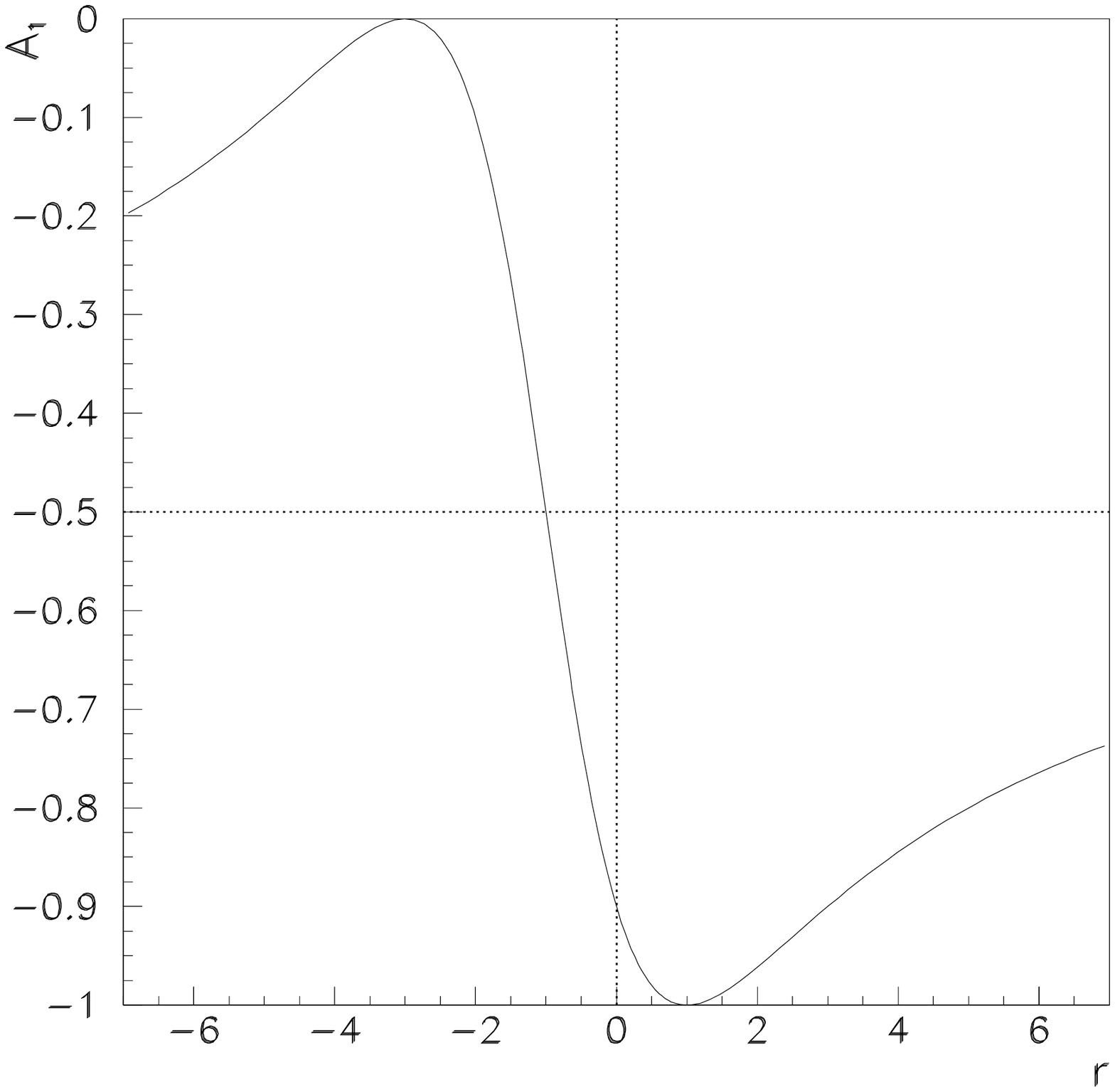}}
\vspace*{.2 truecm}
\caption{Dependence of ${\cal A}_1$ on $r$}.
\label{fig:fig4}

\end{figure}\begin{figure}
\mbox{\epsfysize=15.cm\leavevmode \epsffile{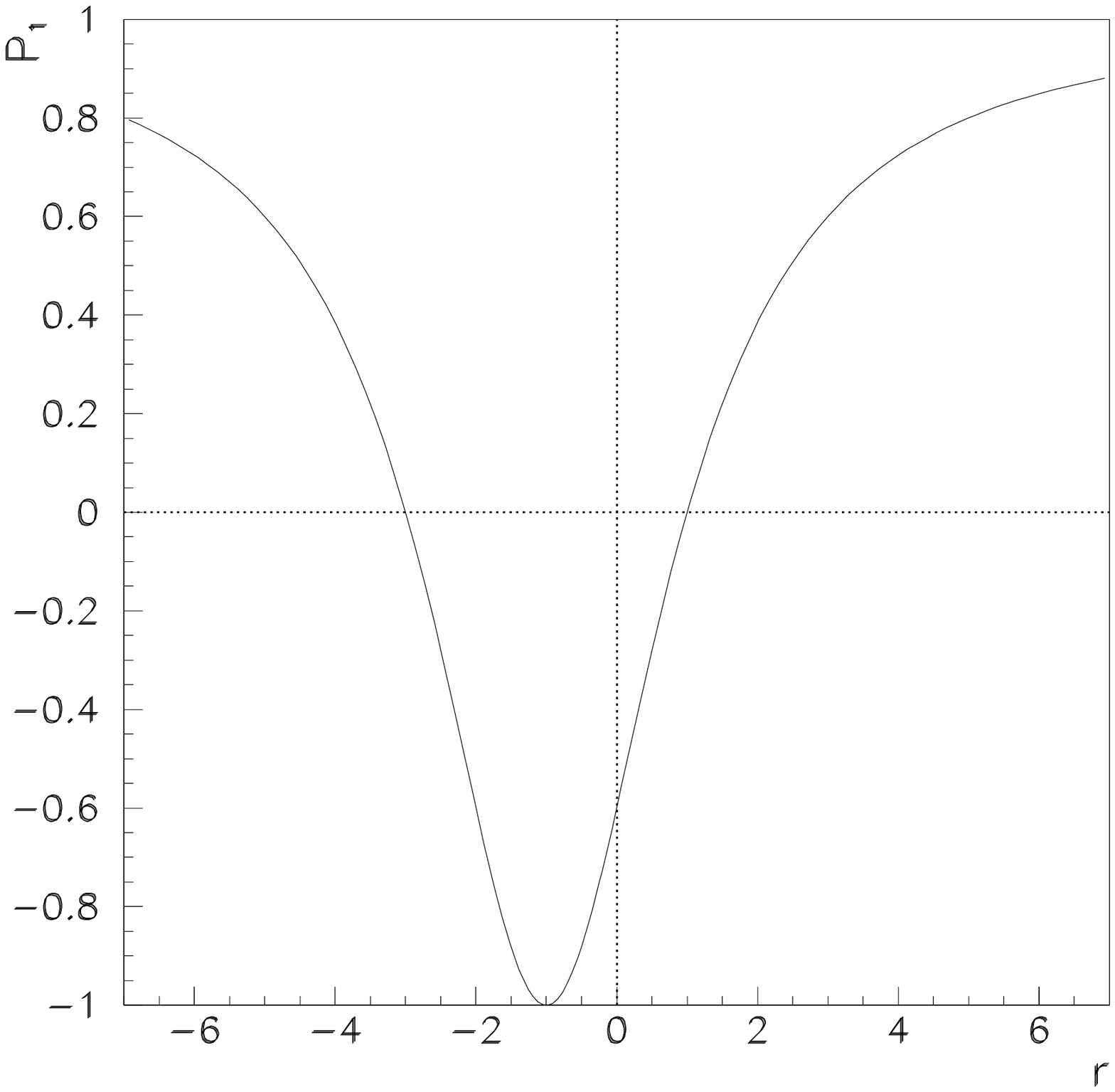}}
\vspace*{.2 truecm}
\caption{Dependence of ${\cal P}_1$  on $r$}.

\label{fig:fig5}
\end{figure}

\begin{figure}
\mbox{\epsfysize=15.cm\leavevmode \epsffile{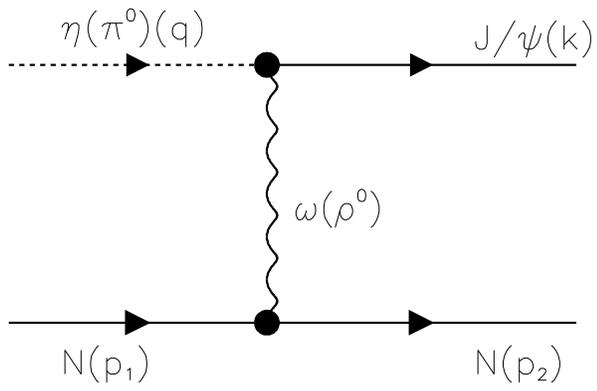}}
\vspace*{.2 truecm}
\caption{Feynman diagram for $t$-channel $\omega(\rho)$-exchange 
for the process $\eta(\pi)+N\to N+N+J/\psi$.}
\label{fig:fig6}
\end{figure}

\end{document}